\documentclass[sn-mathphys, a4, iicol]{sn-jnl}% Math and Physical Sciences Reference Style

\usepackage{newtxtt}
\usepackage{newtxmath, times, float}
\usepackage{xcolor}

\usepackage[utf8]{inputenc}

\usepackage{graphicx}
\graphicspath{{figures/}}
\DeclareGraphicsExtensions{.pdf,.jpeg,.png,.eps}

\newfloat{lstfloat}{htbp}{lop}
\floatname{lstfloat}{Listing}
\usepackage{listings}

\definecolor{commentsColor}{rgb}{0.0, 0.5, 0.0}
\definecolor{keywordsColor}{rgb}{0.25, 0.25, 0.5}
\definecolor{stringColor}{rgb}{0.5, 0.0, 0.1}

\lstMakeShortInline[]|
 
\jyear{2023}%

\theoremstyle{thmstyleone}%
\theoremstyle{thmstyletwo}%

\theoremstyle{thmstylethree}%

\unnumbered % uncomment this for unnumbered level heads

\begin{document}

% \title{Training and Deploying Spiking NN Applications to the Mixed-Signal Neuromorphic Chip DYNAP\texttrademark-SE2 with Rockpool}

\title[~]{Gradient-descent hardware-aware training and deployment for mixed-signal Neuromorphic processors}

%%=============================================================%%
%% Prefix	-> \pfx{Dr}
%% GivenName	-> \fnm{Joergen W.}
%% Particle	-> \spfx{van der} -> surname prefix
%% FamilyName	-> \sur{Ploeg}
%% Suffix	-> \sfx{IV}
%% NatureName	-> \tanm{Poet Laureate} -> Title after name
%% Degrees	-> \dgr{MSc, PhD}
%% \author*[1,2]{\pfx{Dr} \fnm{Joergen W.} \spfx{van der} \sur{Ploeg} \sfx{IV} \tanm{Poet Laureate} 
%%                 \dgr{MSc, PhD}}\email{iauthor@gmail.com}
%%=============================================================%%

\author[1]{\fnm{Ugurcan} \sur{Cakal}}\email{ugurcan.cakal@synsense.ai}

\author[2]{\fnm{Maryada}}\email{maryada@ini.uzh.ch}

\author[1]{\fnm{Chenxi} \sur{Wu}}\email{chenxi.wu@synsense.ai}

\author[3]{\pfx{Prof} \fnm{Ilkay} \sur{Ulusoy}}\email{ilkay@metu.edu.tr}

\author[1]{\pfx{Dr} \fnm{Dylan} \sur{Muir}}\email{dylan.muir@synsense.ai}

\affil[1]{\orgname{SynSense AG}, \orgaddress{\street{Thurgauerstrasse 60}, \city{Z\"urich}, \postcode{8050}, \country{Switzerland}}}

\affil[2]{\orgdiv{Institute of Neuroinformatics}, \orgname{University \& ETH Z\"urich}, \orgaddress{\street{Winterthurerstrasse 190}, \city{Z\"urich}, \postcode{8057}, \country{Switzerland}}}

\affil[3]{\orgdiv{Electrical and Electronics Engineering}, \orgname{METU}, \orgaddress{\city{Ankara}, \postcode{06800}, \country{Turkey}}}

%%==================================%%
%% sample for unstructured abstract %%
%%==================================%%

\abstract{
    Mixed-signal neuromorphic processors provide extremely low-power operation for edge inference workloads, taking advantage of sparse asynchronous computation within Spiking Neural Networks (SNNs).
    However, deploying robust applications to these devices is complicated by limited controllability over analog hardware parameters, as well as unintended parameter and dynamical variations of analog circuits due to fabrication non-idealities.
    Here we demonstrate a novel methodology for offline training and deployment of spiking neural networks (SNNs) to the mixed-signal neuromorphic processor DYNAP-SE2.
    The methodology utilizes gradient-based training using a differentiable simulation of the mixed-signal device, coupled with an unsupervised weight quantization method to optimize the network's parameters.
	Parameter noise injection during training provides robustness to the effects of quantization and device mismatch, making the method a promising candidate for real-world applications under hardware constraints and non-idealities.
    This work extends Rockpool, an open-source deep-learning library for SNNs, with support for accurate simulation of mixed-signal SNN dynamics.
    Our approach simplifies the development and deployment process for the neuromorphic community, making mixed-signal neuromorphic processors more accessible to researchers and developers.
}

\keywords{mixed-signal, neuromorphic, spiking neural networks, DYNAP-SE2}

\maketitle

\section{Introduction}\label{intro}

Neuromorphic processors use analog and mixed-signal circuits to emulate the dynamics and computational abilities of biological neurons and synapses.
One of the most advanced architectures is DYNAP-SE2\cite{Richter2023}, which has an asynchronous mixed-signal structure whose analog components operate in the subthreshold operation range, making it an ultra-low power and ultra-low latency application delivery candidate.

Devices such as DYNAP-SE2 offer a high degree of biological realism and configurability, but have been historically difficult to configure for several reasons: their complex architecture; the vast number of parameters due to high configurability; and the lack of standardized configuration protocols.
Despite this difficulty, DYNAP-SE family members have already been used in several low-dimensional signal processing applications.
In \cite{bauer20}, real-time classification of heartbeat pathologies from multi-channel electrocardiogram (ECG) recordings was performed, distinguishing between nominal beats and pathological rhythms.
In \cite{donati18} and \cite{donati19}, Electromyography (EMG) signals were analysed to distinguish the movement of hand muscles to classify gestures.
In these applications, the reservoir computing paradigm \cite{maas02} was exploited.
A semi-randomly initialized spiking recurrent neural network (SRNN) was deployed to the DYNAP-SE device to integrate the temporal patterns hidden in sensory signals.
Classification was performed by a linear readout implemented on a conventional CPU, by monitoring the spiking activities of hardware neurons on DYNAP-SE.
These applications demonstrated that RSNN inference on the Dynap-SE chip can operate in the sub-mW power range.

DYNAP-SE2 and other similar devices individually instantiate arrays of synapses and neurons in analog circuits.
Analog nature of these emulation circuits exposes them to variability of the individual device components, due to variations introduced in the fabrication process.
This variability, known as "mismatch", introduces a unique diversity in the behavior of ostensibly identical neurons and synapses across and between chips, enhancing adaptability~\cite{buechel21, Zendrikov2022}.
Mixed-signal devices such as DYNAP-SE2 do not usually permit individual control over each parameter on the chip, instead grouping parameters such as time-constants, thresholds, and even weight values across a number of neurons and synapses.
This grouping reduces the parameter space of the device. 
While it inherently presents a challenge in calibration, the additional factor of mismatch further intensifies this challenge, simultaneously opening avenues for advanced calibration techniques and innovation.
In addition, grouping parameters means that the parameter configuration space of an SNN deployed to the chip is itself heavily constrained.

Training and deploying applications to these devices involves a thorough and detailed process, requiring several months of dedicated effort from skilled researchers. 
To address this challenge, our work introduces an efficient and effective methodology that provides the potential for commercial application development for DYNAP-SE2.
Extending Rockpool, an open-source deep-learning library for SNNs~\cite{muir19}, our toolchain performs offline gradient-based hardware-aware optimization of SNNs, which can be robustly deployed at scale to mixed-signal devices, while preserving behavior. 
This Rockpool tutorial, available at \href{https://rockpool.ai/devices/DynapSE/jax-training.html}{https://rockpool.ai/devices/DynapSE/jax-training.html} offers an in-depth Jupyter Notebook on how to train a spiking network for use with the DYNAP-SE2 processor.
It reproduces the experiment introduced in this paper, and comprehensively covers creating synthetic datasets, constructing and fine-tuning a spiking neural network with Rockpool and Jax, and evaluating the network's performance.
Additionally, it addresses gradient-based optimization techniques and the complexities of device mismatch in mixed-signal chips.

This work provides a DYNAP-SE2 simulator, ``DynapSim'', which operates in the same parameter space as DYNAP family processors~\cite{moradi18},~\cite{Richter2023}.
DynapSim executes an efficient and accurate differentiable simulation of the DYNAP-SE2 design dynamics to solve the characteristic circuit transfer functions over time.
This \textit{differentiable computing} approach has in recent years been applied to SNNs to train deep spiking networks with gradient-based methods borrowed from machine learning~\cite{Neftci2019, Eshraghian2023}.

In order to perform gradient-based optimization, a spiking neuron model requires an additional surrogate gradient function to ensure loss gradients can propagate through the neuron~\cite{Lee2016, Zenke2018, Neftci2019, Kaiser2020}.
Broadly speaking, this addresses the issue that the derivative of the spike generation functions used in spiking neurons are ill-formed, resulting in zero or undefined gradients when propagating through the neuron.
By providing a \textit{surrogate gradient} for the spike generation function, loss gradients can be preserved.
In our DYNAP-SE2 neuron implementation, taking the derivative of the output spike train $S_{out}(t)$ with respect to a parameter $P$ that affects the membrane current dynamics appears as follows:

$$
    \dfrac{\partial S_{out}(t)}{\partial P} = \dfrac{\partial \Theta (I_{mem}, I_{spkthr})}{\partial I_{mem}} \cdot \dfrac{\partial I_{mem}}{\partial P}
$$

Here, $\Theta(\cdot)$ denotes the Heaviside step function, and $P$ is any parameter that changes the membrane dynamics such as a leakage current, gain current, etc.
In order to determine the effect of changing the parameter $P$ on the output spike train, ${\partial \Theta(\cdot)}/{\partial I_{mem}}$ must be well defined.
However, the derivative of $\Theta(\cdot)$ is zero everywhere and infinte at the spiking threshold.
As a solution to this problem, an continuous surrogate function is defined that substitutes the derivative of $\Theta(\cdot)$ in the backward pass of the backpropagation algorithm.
In particular, we adopt a rectified linear function (ReLU) as a surrogate, with constant derivative when $I_{mem} > I_{reset}$.
For further implementation details of the neuron model and the surrogate function see the Rockpool tutorial \href{https://rockpool.ai/devices/DynapSE/neuron-model.html}{https://rockpool.ai/devices/DynapSE/neuron-model.html}

DynapSim is used as a computational neuron model in offline SNN simulations and during training.
Rockpool translates the optimized networks to equivalent hardware configurations, and provides straightforward deployment of these networks to DYNAP-SE2 chips.
For an overview of the DYNAP-SE2 hardware, one can refer to \cite{Richter2023}.
We then describe the implementation details of DynapSim, and demonstrate training, deployment and quantitative evaluation of a toy model to DYNAP-SE2 hardware.
We present the steps to achieve training and deployment using code examples for Rockpool.

Our implementation is available as part of the open-source Python package Rockpool: \href{https://rockpool.ai/devices/DynapSE/dynapse-overview.html}, with code available at \href{https://github.com/synsense/rockpool}.

\subsection{Overview of the DYNAP-SE2 Hardware}\label{mixed-signal}

\begin{figure}[t]
    \centering
    \includegraphics[width=\linewidth]{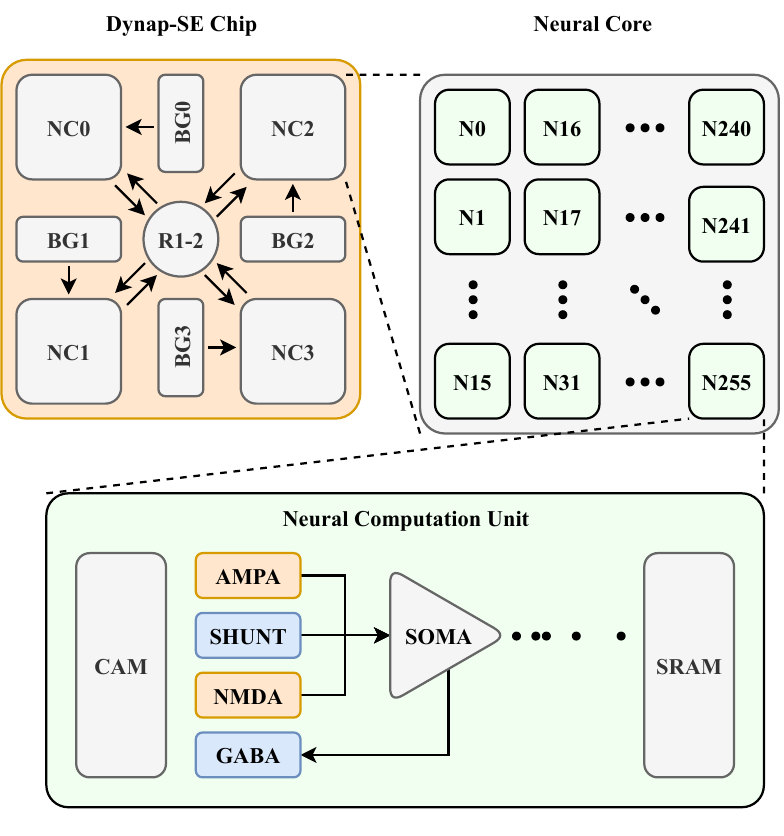}
    \caption{
        \textbf{DYNAP-SE2 Architecture.}
        See text for details.
        NC: Neural Core; R: Router; BG: Bias Generator. Other acronyms: see text.
    }
    \label{fig:dynapse_architecture}
\end{figure}

The DYnamic Neuromorphic Asynchronous Processor --- ScalablE 2 (DYNAP-SE2) is a mixed-signal chip that inherits the event-driven nature of the DYNAP family~\cite{moradi18, Richter2023}.
It directly emulates biological behavior using analog spiking neurons and analog synapses as the computational units.
The transistors of the neural cores operate in the subthreshold regime, resulting in power consumption below 1~mW.
%, which is about one-thousandth to one-millionth of the state-of-the-art digital neuromorphic chips.
Each DYNAP-SE2 chip is equipped with 1024~adaptive exponential integrate-and-fire (AdExpIF) analog ultra-low-power spiking neurons and 64~synapses per neuron.
Fig.~\ref{fig:dynapse_architecture} includes an overview of the architecture of the chip.

The DYNAP-SE2 digital spike routing architecture involves pre-synaptic neurons broadcasting their spiking activities on an internal bus using specific neuron ``tags''.
Post-synaptic neurons monitor the bus for up to 64 tags each, such that they can connect to up to 64 pre-synaptic tag IDs.
This approach enables both sparse and dense connection patterns, and since multiple neurons may broadcast the same tag, permits a fan in of greater than 64 pre-synaptic neurons.

In this routing system, post-synaptic neurons themselves effectively store the weight matrices through their broadcasting and listening connections.
The strength of each synaptic weight is determined by the weight current configuration stored at each synapse.
This means that the synaptic efficiency, or the impact of a pre-synaptic neuron's signal on a post-synaptic neuron, is determined by the amount of current assigned to that particular synaptic connection.
For more comprehensive information on the specific mechanisms of weight storage and transfer in this architecture, refer to \cite{Richter2023}, and \cite{cakal22}.

The neural computation unit serves as the primary building block for creating the dynamics in DYNAP-SE2.
Each neural core (NC) consists of 256~analog neurons that share the same parameter set.
The digital memory blocks, Content-Addressable Memories (CAMs) and Static-RAMs (SRAMs), store the transmitting and receiving event configurations, respectively.
The synapses and neuron soma carry out analog computations, with four different types of synapses --- AMPA, GABA, NMDA, and SHUNT --- integrating the incoming events and injecting current into the membrane.
AMPA and NMDA activation increase the firing probability, while GABA and SHUNT activation decrease it.

The CAM stores the listening event setting for each of the 64 connections of a neuron, which specifies its synaptic processing unit.
The neuron soma integrates the injection currents and holds a temporal state, with configurable paths of charging and discharging capacitors designating the temporal behavior.
The membrane current, which is a secondary reading on the membrane capacitance, functions as the temporal state variable.
When the membrane current in a neuron reaches the firing threshold a reset mechanism is activated, which returns the neuron membrane potential to a reset state. 
This also triggers the event sensing units, which then package the event in Address Event Representation (AER) format and broadcast it on the internal bus.
In this way, the neuron uses analog sub-threshold circuits to compute the dynamics but conveys the resulting outputs using a digital routing mechanism.

Each neural core holds a parameter group to set the neuronal and synaptic parameters for its 256~neurons and their pre-synaptic synapses.
The neurons in the same core share the same parameter values, including time constants, refractory periods, synaptic connection strengths, and other attributes.
Special digital-to-analog converters, bias generators (BG), set these parameter current values.
In total, there are 70~parameters that can be set to adjust the behavior of the neurons and synapses, including time constants, pulse widths, amplifier gain ratios, and synaptic weight strengths, among others.
A comprehensive table detailing these parameters, and their impact on the SNN simulation is provided in the appendices for reference, Table~\ref{tab:all_params}.

For simulation purposes, a custom computational spiking neural model relates the behavioral dynamics of a computational neural setting to the VLSI parameters of the respective circuits.
It uses forward Euler updates to predict the time-dependent dynamics and solves the characteristic circuit transfer functions in time.
Specifically, a ``DynapSim'' neuron solves the silicon neuron~\cite{livi09} and silicon synapse~\cite{bartolozzi07} circuit equations, making use of assumptions and simplifications from~\cite{chicca14}.
Further details of the application and implementation can be found in~\cite{cakal22}.

\section{Methods}

\subsection{Task and Training approach}\label{training}

DynapSim is an extension of the contemporary spiking neural network library, Rockpool~\cite{muir19}, and serves as a simulation solution for the device.
The solution it offers involves solving characteristic equations of the analog circuits and does not provide a circuit-level accurate simulation.
Instead, DynapSim provides an approximate simulation that can be fine-tuned and translated into a device configuration.
The simulator is powered by the state-of-the-art high-performance machine learning library JAX~\cite{jax2018github}, which facilitates fast execution and just-in-time compilation on CPUs, GPUs and TPUs.
The toolchain we provide performs off-chip gradient-based optimization of an SNN and deploys the trained network to the chip while preserving the optimized behavior.
The upcoming sections elaborate on the technique of gradient-based optimization employed to train a Spiking Neural Network (SNN) before its deployment to a DYNAP-SE2 chip.

\paragraph{Toy task: Frozen Noise Classification}

\begin{figure}[t]
    \centering
    \includegraphics[width=\linewidth]{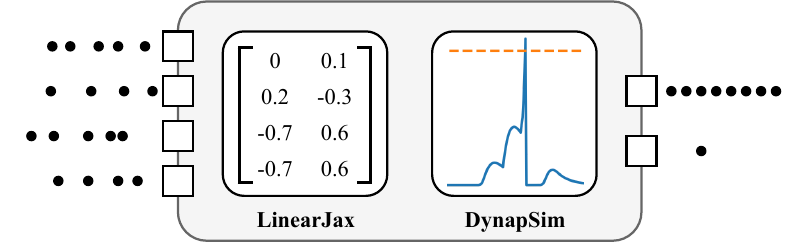}
    \caption{
        \textbf{Frozen noise classification task.}
        60 input channels provide input spiking patterns to the network (left; 4 shown here).
        The network provides two output channels (right), which should emit high spiking activity when presented with one of two target frozen noise inputs.
        ``LinearJax'' and ``DynapSim'' modules, provided by Rockpool, are used to simulate the weights, synapse and neuron dynamics of the network (see text for further details).
    }
    \label{fig:frozen_noise}
\end{figure}

The purpose of the frozen noise classification experiment is to evaluate the learning abilities of the implemented simulator.
The experiment focuses on training a DynapSim network to accurately classify two distinct random frozen noise patterns.
The network comprises two analog neurons with recurrent connections, along with 60~external input connections.
The desired outcome is for the first neuron to exhibit a significantly higher firing rate when presented with the first frozen noise, and for the second neuron to exhibit a significantly higher firing rate when presented with the second frozen noise.
Fig.~\ref{fig:frozen_noise} illustrates the task at hand.

\paragraph{Data}

To run the experiment with the spiking neuron model, a spiking input pattern is necessary.
For this specific task, randomly generated discrete Poisson time series with a mean frequency of 50~Hz in a 500~ms duration are used as frozen noise recordings.
Each sample comprises 60~channels, and the time-step duration is 1~ms.

For training purposes, two samples are utilized to enable the network to overfit, while 1000~different random samples are reserved for testing the trained network.
The optimized network should recognize the 2~critical training samples, by generating high activity on the corresponding output neuron, and low activity on the other neuron.
If any other sample is provided, the network should generate random output activity.

\begin{figure}[t]
    \centering
    \includegraphics[width=0.8\linewidth]{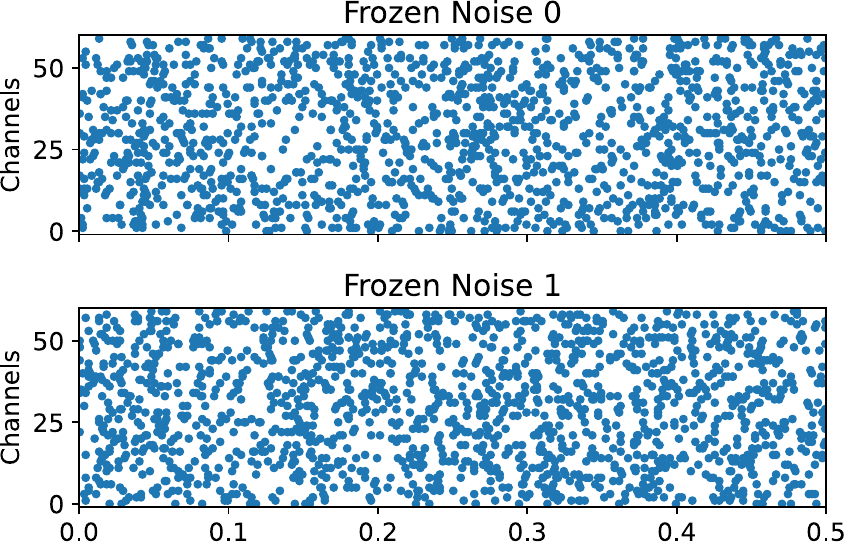}
    \caption{\textbf{Frozen noise recordings used in training.}}
    \label{fig:noise_patterns}
\end{figure}

\paragraph{Network}

The network architecture used in this study consists of a simple recurrent spiking network with weighted inputs.
In Rockpool, this network is constructed of two layers |LinearJax| and |DynapSim|.
The first module, |LinearJax|, applies a linear transformation to the input spikes to simulate spike weighting.
The second module, |DynapSim|, simulates the time-dependent analog silicon neuron and synapse dynamics.
Listing~\ref{lst:network} shows how to instantiate this network in Rockpool.

\begin{lstfloat}
\begin{lstlisting}[language=Python, caption={\textbf{Constructing the SNN in Rockpool.}}, label={lst:network}]
    net = Sequential(
        LinearJax((Nin, Nrec)),
        DynapSim((Nrec, Nrec), dt=dt),
    )
\end{lstlisting}
\end{lstfloat}

This network architecture can be compared to using a ReLU activation layer following a fully connected layer in classical NNs.
The difference, however, lies in the fact that the DynapSim layer computes and maintains a time-dependent state instead of a stateless activation.
The output of the DynapSim neurons depends not only on the instantaneous inputs but also on past inputs via internal state variables.
The state continues to evolve continuously over time, regardless of when the neuron receives spikes on its input.
Additionally, the DynapSim layer encapsulates a recurrent connection matrix that is one of the targets of the optimizer.
Lastly, the layer corresponds to a custom analog hardware configuration, and solving the characteristic equations of the analog circuits is a key part of its function.

To limit the complexity of the task, in this case only the weight parameters are trained, while the rest of the neuron and synapse parameters are fixed to their default simulation values.
This means that mathematically, only two 2D weight matrices are subject to optimization: the $60\times 2$ input weight matrix stored inside LinearJax and the $2\times 2$ recurrent weight matrix stored inside DynapSim.

\paragraph{Response Analysis}\label{sec:response_analysis}
For the frozen noise discrimination task, the classification of the network is indicated by the neuron with the highest mean firing rate $r_0$ for class~0, and $r_1$ for class~1.
As a performance metric, the ratio between the output neurons' mean firing rates quantifies the network's ability to distinguish the two target frozen noise patterns.
The firing rate ratio (FRR) is calculated by dividing the higher mean firing rate by the lower mean firing rate read from the decision neurons, as shown in Eq.~\ref{eq:frr}.
\begin{equation}
    FRR = \dfrac{\max(r_0, r_1)}{\min(r_0, r_1)}
    \label{eq:frr}
\end{equation}

\paragraph{Mismatch Simulation}

\begin{figure}[t]
    \centering
    \includegraphics[width=0.8\linewidth]{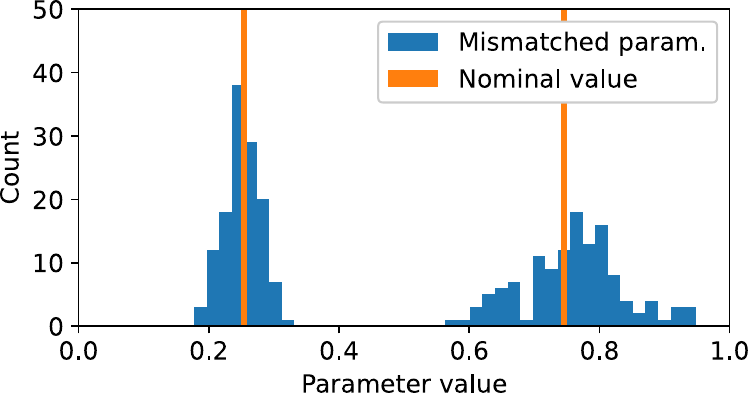}
    \caption{\textbf{Mismatch Simulation.} Nominal values for two parameters (orange) are applied to the network, following which DynapSim is used to simulate the parameter mismatch that would be experienced when deployed to a DYNAP-SE2 device (blue distributions).}
    \label{fig:mismatch_illustration}
\end{figure}

Each optimisation step during training includes a forward and a backward pass.
The forward pass simulates the mixed-signal circuit behavior under ideal conditions, but the circuits in real sub-threshold computation devices are subject to parameter mismatch.
In order to make trained networks robust against parameter deviations, B\"uchel et al. proposed to modify values during training by injecting parameter noise as well as by an adversarial attack on parameter values~\cite{buechel21}.

DynapSim includes an empirically-verified model of parameter mismatch, which we apply in the forward pass, slightly perturbing the parameter values in the network that are subject to change.
This mismatch simulation model addresses parameter variability without focusing on certain known factors like temperature fluctuations or process impurities. 
Instead, it offers a general approach to managing parameter mismatches.
It deviates the parameters using a Gaussian distribution, with the mean taken as the nominal parameter values, and variation determined by empirical measurement~\cite{buechel21, Zendrikov2022}.
New mismatched parameters are set every $n$~epochs.
During optimisation, the network reaches parameter values that obtain a low loss value, in spite of the parameter variation.
As a result, SNNs trained in this way are less sensitive to mismatch-induced loss of performance when deployed to mixed-signal neuromorphic hardware.
Fig.~\ref{fig:mismatch_illustration} illustrates the effect of mismatch on parameter values.

\paragraph{Optimization}

\begin{figure}[t]
    \centering
    \includegraphics[width=0.8\linewidth]{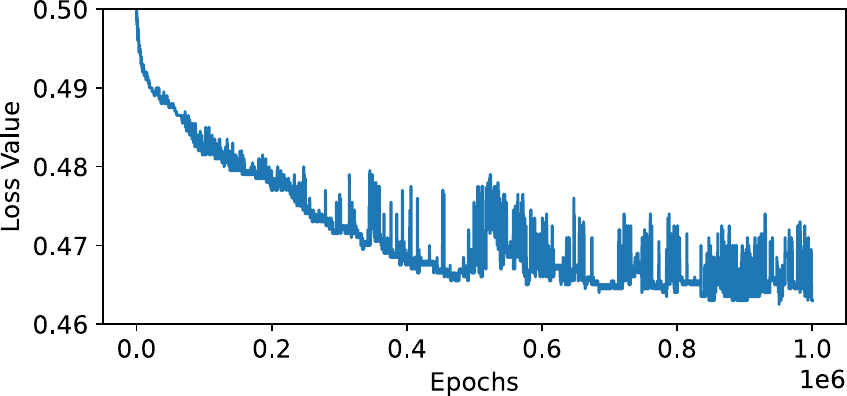}
    \caption{\textbf{Mean square error (MSE) loss over the course of training.}}
    \label{fig:mse_loss}
\end{figure}

The optimization objective is highly dependent on the task and can be customized according to the requirements.
In this particular task, the goal is to increase the firing rate of a specific neuron upon receiving a known frozen noise record.
To achieve this, the mean square error (MSE) loss function is utilized.

The target signal is a uniform spike train that generates an event at every time step from one channel and that generates no events from the other channel.
The mean value of the differences in rate between target and network output gives a scalar loss value to be used in error backpropagation.

In this experiment, the Adam algorithm is utilized for optimization~\cite{kingma15}, and the training pipeline is similar to a conventional machine learning pipeline.
Since the forward computation involves non-differentiable spike production functions, a surrogate gradient approximation replaces these in the backward pass~\cite{Lee2016, Zenke2018, Neftci2019, Kaiser2020}.

\begin{figure*}[t]
    \centering
    \includegraphics[width=0.7\linewidth]{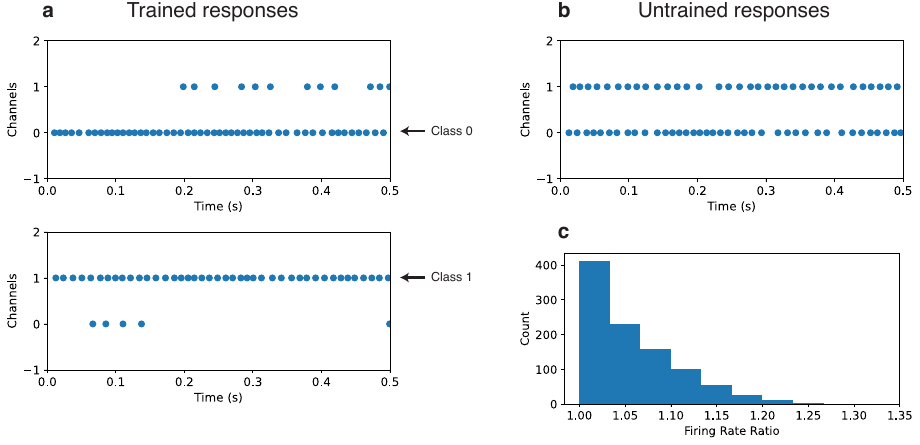}
    \caption{
        \textbf{Response of the simulated trained network to trained and untrained input samples.}
        \textbf{a} Response of the trained network to input classes~0 and~1. FRRs 4.1 and 8.4, respectively for these samples.
        \textbf{b} Response of the trained network to an untrained Poisson noise sample.
        \textbf{c} Distribution of FRR of the trained network to 1000 untrained noise samples. FRRs are close to 1.0, indicating the network has learned to reject these unknown inputs.
    }
    \label{fig:trained_net_response}
\end{figure*}

Fig.~\ref{fig:mse_loss} illustrates the decrease in MSE loss over the training process.
During training, the MSE loss decreased from 0.5 to 0.46 over one million epochs. 
Despite the straightforward task, a small learning rate was necessary to provide stable learning in the face of the complex non-linear neuron model.
Even this seemingly small reduction in loss value results in a significant improvement in behavior, allowing the network to classify two similar frozen noise samples.
Fig.~\ref{fig:trained_net_response}a shows the response of the network to trained input samples.

When the first noise pattern is presented to the network, neuron~0 (channel~0) exhibits high firing activity (164~Hz), while neuron~1 (channel~1) shows significantly lower activity (24~Hz), resulting in an FRR of 6.83.
On the other hand, when the second noise pattern is presented, neuron~1 (channel~1) fires almost constantly (122~Hz), while neuron~0 (channel~0) remains quiet as intended (10~Hz), resulting in an FRR of 12.2.
The clear distinction between the higher and lower firing rates demonstrates that the network is capable of distinguishing between the two input patterns.

To evaluate the network's recognition capabilities on unseen data, we used a test set of random noise samples to demonstrate that the network only recognizes the training patterns.
We generated 1000 frozen noise input patterns with the same mean frequency and length as the target patterns used in training.
The FRRs between the decision neurons were recorded to quantify the ability of the network to reject un-trained input patterns.
We expect the network to respond with FRRs close to one, indicating the input patterns are not similar to either class~0 or class~1.
Fig.~\ref{fig:trained_net_response}b shows the network's repsonse to a random sample, with an FRR close to 1.
The distribution of FRR values under 1000 random noise samples is shown in Fig.~\ref{fig:trained_net_response}c.

During the 500~ms test runs, both the first and second neurons remain active, firing at similar rates.
Based on these observations, it can be concluded that the network responds strongly only to the trained target inputs, and rejects the non-trained noise inputs.

\subsection{Deployment to DYNAP™-SE2}\label{mapping-deployment}

To deploy an SNN using Rockpool, the network is first defined in simulation and optimized using gradient-based or non-gradient-based methods.
Rockpool then extracts a computational graph from the optimized network, containing all the necessary parameters for specifying the chip configuration.
However, the computational graph does not include information about hardware resource allocation, so a mapping procedure is required to cluster the parameters and find a suitable hardware allocation.
The parameters also need to be quantized since the DYNAP-SE2 hardware cannot support floating point precision, and converted to bias values to configure the neuron and synapse parameters.
Finally, the user needs to connect and interface with the chip to deploy the SNN.

Rockpool accomplishes this process in only a few lines of code, as demonstrated in Listing~\ref{lst:map}.

\begin{lstfloat}[t]
\begin{lstlisting}[language=Python, caption={\textbf{Deploying an SNN to DYNAP-SE2.}}, label={lst:map}]
# Define
net = Sequential(
    LinearJax((Nin, Nrec)),
    DynapSim((Nrec, Nrec), dt=dt),
)

# Map
spec = mapper(net.as_graph())
spec.update(autoencoder_quantization(**spec))
config = config_from_specification(**spec)

# Connect & Interface
se2_devices = find_dynapse_boards()
se2 = DynapseSamna(se2_devices[0], **config)
out, state, rec = se2(raster, record=True)
\end{lstlisting}
\end{lstfloat}

Our pipeline also supports ``reverse mapping'', whereby a simulation SNN can be extracted from an existing hardware configuration, as shown in Listing~\ref{lst:hardware_snn}.

\begin{lstfloat}[t]
\begin{lstlisting}[language=Python, caption={\textbf{Extracting an SNN from a hardware configuration.}}, label={lst:hardware_snn}]
    net = dynapsim_net_from_config(**config)
    out, state, rec = net(raster, record=True)
\end{lstlisting}
\end{lstfloat}

The sections below explain the details of these steps. 

\paragraph{Computational Graph}

A computational graph in Rockpool represents the flow of data through a neural network.
In Rockpool, the $|as_graph()|$ method extracts a computational graph from a full SNN model.
This graph captures all the computationally significant parameters of the network, as well as the network structure.
The graph representation enables manipulation of the SNN architecture, facilitating mapping the network parameters to various hardware architectures, and permitting conversion between neuron models.
For instance, a trained LIF network can be transformed into an equivalent DynapSim network, with similar behaviour.

\paragraph{Mapping}
The |mapper()| functionality converts a computational graph from an arbitrary SNN into a DYNAP-SE2 HDK hardware specification, regardless of whether the network was originally a DynapSim network.
|mapper()| clusters the parameters into groups and determines the hardware IDs of neurons.
The mapping process is described in detail below.
For an alternative approach to mapping SNNs to DYNAP-SE hardware, see\cite{balaji20}.

\begin{figure*}[t]
    \centering
    \includegraphics[width=0.8\linewidth]{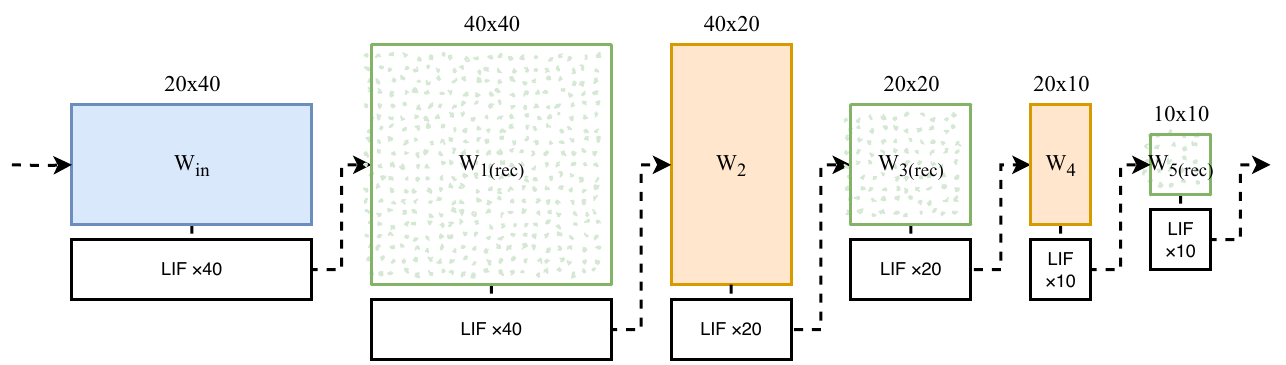}
    \caption{
        \textbf{Weight matrix parameters and information flow through an example SNN.}
        Input weights $W_\textrm{in}$ are shown in blue.
        Blocks of recurrent weights $W_{n(\textrm{rec})}$ are hatched.
        Feed-forward hidden weights $W_n$ are indicated in orange.
        Blocks of LIF spiking neurons are indicated in black.
    }
    \label{fig:snn_open}
\end{figure*}

Fig.~\ref{fig:snn_open} shows the weight parameters and information flow in an example SNN with both feed-forward and recurrent components. 

\begin{figure}[t]
    \centering
    \includegraphics[width=\linewidth]{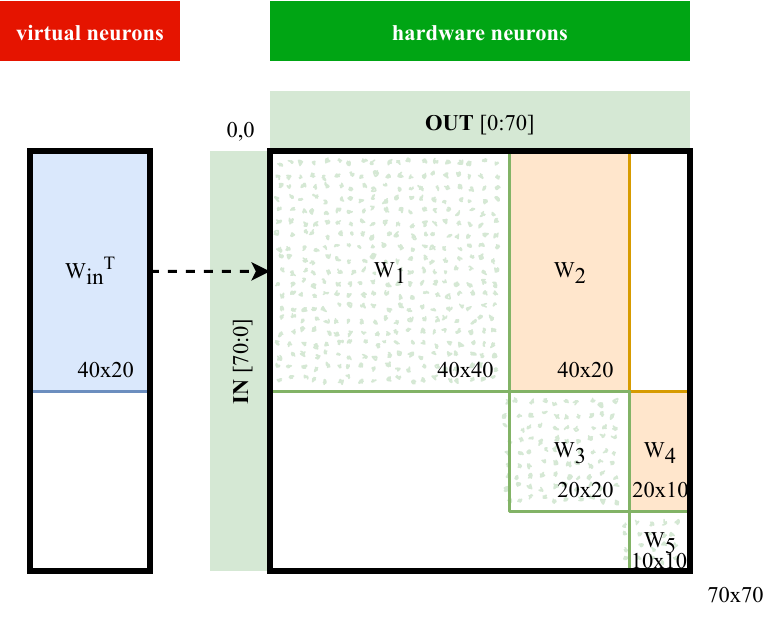}
    \caption{
        \textbf{Merged weight matrices given the network in Fig.~\ref{fig:snn_open}.}
        The blocks of weights in Fig.~\ref{fig:snn_open} are indicated here with their corresponding colours and labels.
    }
    \label{fig:snn_closed}
\end{figure}

The input weight matrix, $W_{in}$, applies a linear transformation to the external input and delivers it to the hardware neurons. 
Recurrent weight matrices, $W_{1(rec)}$, $W_{3(rec)}$, and $W_{5(rec)}$, establish the connection weights between hardware neurons, while feed-forward weight matrices, $W_{2}$ and $W_{4}$, connect different groups of neurons to each other. 
The mapper produces a single equivalent recurrent weight matrix that collects all feed-forward neurons in a single pseudo-recurrent representation.

Fig.~\ref{fig:snn_closed} shows the merged input and recurrent weight matrices corresponding to the network in Fig.~\ref{fig:snn_open}.

The input weight matrix $W_{in}$ connects virtual input neurons to hardware neurons on DYNAP-SE2
All other weight matrices in Fig.~\ref{fig:snn_open} are merged into one large recurrent weight matrix. 
The input neurons are assigned tags (virtual IDs) from the set of virtual input tags (``virtual tags''), while the hardware neurons are assigned tags (hardware IDs) from the list of available hardware neurons (``actual tags'').

Once this is complete, the |mapper()| reduces an SNN down to three connected graph modules: one |DynapseNeurons| object holding the current parameter values of the hardware neurons, one |LinearWeights| object holding the input weights from the external connections to the hardware neurons, and one |LinearWeights| object holding the recurrent weights between the hardware neurons.

\paragraph{Quantization}
During simulation, weight matrices in the layers can adopt any floating-point value, but when deploying to the hardware of DYNAP-SE2, weight settings are limited to a 4-bit restricted connection-specific assignment. 
To convert weight matrices to device configuration, a quantization phase is necessary. 
The quantization process for DYNAP-SE2 involves two steps: defining 4 base weight parameters for the inner product space and storing connection-specific 4-bit binary weight masks in digital memory cells.
The goal of quantization is to find a set of ``base weight'' parameter values and a binary bit-mask matrix that together reconstruct a floating-point weight matrix with minimal deviation.
To accomplish this, an auto-encoder structure, a popular unsupervised machine learning method, is used.
The intermediate code representation represents the base weight currents, and the decoder weight matrix provides binary bit-masks.

\begin{figure}[t]
    \centering
    \includegraphics[width=\linewidth]{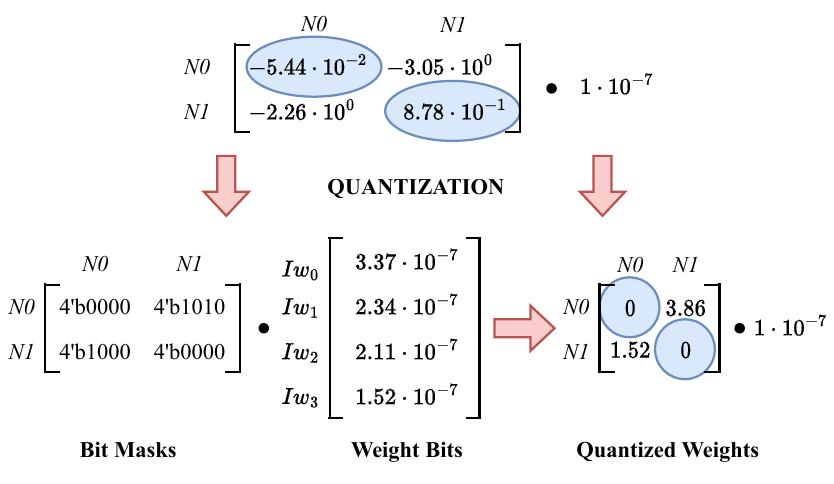}
    \caption{
        \textbf{Process of weight quantization.}
        The floating-point weight matrix (top) is converted into a quantized representation, consisting of a bit-mask containing the number of quantal connections made between two neurons as well as connection type (bottom; left) and a set of ``base weight'' parameters (bottom; middle) indicating the strength of each quantal connection.
        The reconstructed weight matrix (bottom; right) will be similar to, but not identical to the original matrix.
        For example, weak connections may be pruned (blue highlights).
    }
    \label{fig:weight_quantization}
\end{figure}

In the simulated network, weight values can be positive or negative, representing the synapse's excitatory or inhibitory behavior.
The sign of each weight determines the synapse type for that connection: inhibitory GABA synapses for negative values and excitatory AMPA synapses for positive values.
Fig.~\ref{fig:weight_quantization} shows the weight quantization procedure.

Training of the unsupervised auto-encoder learns a hardware-compatible configuration that replicates the target weight matrix with minimal deviation. 
The Mean-Squared Error (MSE) loss metric between the target and reconstructed weight matrices is optimised during the quantization process.

\paragraph{Deployment}
The behavior of a neuron and synapse is characterized by several parameters, including time constants that determine leakage rate and gain ratios that control the amplitude of spike-dependent jumps.
While in simulation these parameters can be adjusted mathematically to modify the behaviour of neurons, implementing this parameterisation in VLSI circuits is more complicated.
Silicon-based implementations of neurons and synapses rely on adjusting bias voltages and currents.
Deploying an SNN application to a mixed-signal device implies translating simulation parameters and the behavioral dynamics of a computational neuron model, into the parameters and bias voltages of the corresponding neuron circuits.
This involves finding a digital bias generator setting that accurately expresses bias current values in Amperes using empirical lookup tables.

\begin{figure}[t]
    \centering
    \includegraphics[width=\linewidth]{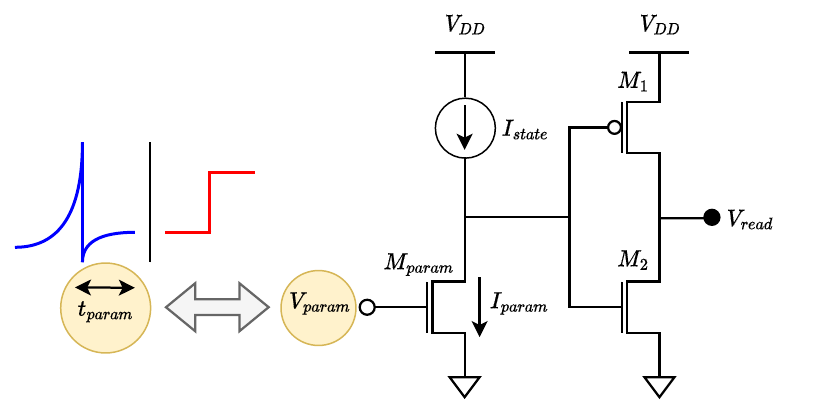}
    \caption{
        \textbf{Translation between high-level behavioural parameters and circuit bias values.}
        The ideal behaviour of a neuron (e.g. a time constant; left) corresponds to setting an accurate analog bias value to control the behaviour of a circuit (right).
    }
    \label{fig:simulator}
\end{figure}

Fig.~\ref{fig:simulator} illustrates what parameter translation entails within this reference frame. 
The mapping from high-level parameters to hardware configuration involves two major steps. 
The first step is to identify a supporting current value in amperes, given the parameter. 
For instance, to ensure $\tau = 1 ms$, $I_{leak}$ needs to be $800 pA$ for the AMPA gate, according to our theoretical expectations using the Equation \ref{eq:tau_leak}

\begin{equation}
    \tau = \dfrac{C_{syn}U_{T}}{\kappa I_{\tau}}
    \label{eq:tau_leak}
\end{equation}

Here, $C_{syn}$ represents the synaptic capacitance, $\kappa$ denotes the mean subthreshold factor (n-type, p-type), and $U_T$ represents the thermal voltage, which is approximately 25 mV at room temperature. 
It should be noted, however, that to maintain the simplicity and efficiency of the simulation, the effect of temperature variation due to circuit power dissipation has not been considered.

The second step involves identifying a suitable bias generator configuration, which yields the exact current value needed to set the high-level parameter. To do this, we utilized empirical recordings of bias current responses—given the digital bias generator configuration—as a guide in the digital configuration search. 

While the process explained here applies to all configuration parameters, it does so with a different first step. 
For weight matrices, the weight quantization process returns the base weight currents, and amplifier gains are subsequently computed relative to the leakage currents, depending on predefined ratios.
For a more detailed analysis of the conversion methodology, please refer to \cite{cakal22}.

\section{Results}\label{results}

We used Rockpool and DynapSim to train an SNN as described above, then used the mapping quantization and deployment facilities of Rockpool to deploy the trained SNN to the DYNAP-SE2 chip, configuring the hardware parameters on the chip.
We converted the frozen noise patterns to real-time AER sequences for injecting into a DYNAP-SE2 device.
Each event in the noise pattern is encapsulated with its time and address and sent to the development kit, where an on-board FPGA circuit converts the AER events to digital pulses that stimulate the synaptic input gates of the neurons \cite{Richter2023}.
The analog neurons on the DYNAP-SE2 then process the inputs and produce output events.
Whenever a neuron fires, digital circuits on the FPGA capture the event's timestamp and source address and encapsulate it as an AER event, which is temporarily stored in buffers implemented inside the FPGA.
The output of the hardware evaluation of an input is recorded as AER event sequences.

\begin{figure}[t]
    \centering
    \includegraphics[width=0.8\linewidth]{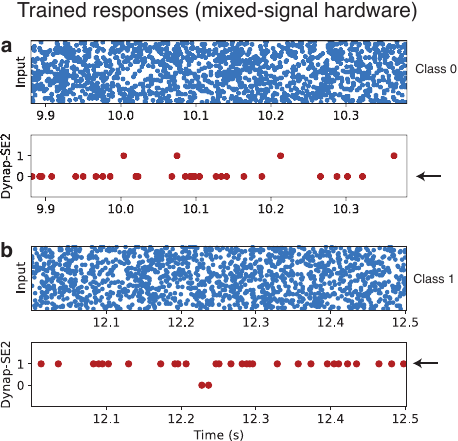}
    \caption{
        \textbf{Response of the trained network deployed to DYNAP-SE2.}
        \textbf{a} Frozen noise for input class~0 (blue) and response of HW neurons (red).
        Arrow: target class.
        \textbf{b} Response to class~1 input.
    }
    \label{fig:hardware_response}
\end{figure}

Fig.~\ref{fig:hardware_response} demonstrates that the combined effects of quantization and device mismatch do not result in loss of performance.
The hidden temporal information that the network has learned is still present, and the network is able to differentiate between the training samples.

The firing rate ratio (FRR) described in the Response Analysis section under Methods is utilized to assess the neurons' ability to distinguish frozen noise patterns. 
A high FRR (>1) indicates good discrimination between the two trained frozen noise input samples.
Tab.~\ref{tab:fn_rate_response} displays the firing responses of the simulated, quantized, and hardware networks when presented with trained frozen noise~0 (FN0), trained frozen noise~1 (FN1), and the un-trained random noise TEST samples.

\begin{table*}[t]
    \centering
    \footnotesize
    \begin{tabular}{@{}lrrlrrlrrl@{}}
        \toprule
        \multirow{2}{*}{\textbf{Frozen Noise}} & \multicolumn{3}{c}{\textbf{Simulated}} & \multicolumn{3}{c}{\textbf{Quantized}} & \multicolumn{3}{c}{\textbf{Hardware}}                                                                                       \\
        ~                                      & \textbf{N0 (Hz)}                            & \textbf{N1 (Hz)}                            & \textbf{FRR}\hspace{3ex}                          & \textbf{N0 (Hz)} & \textbf{N1 (Hz)} & \textbf{FRR}\hspace{3ex}  & \textbf{N0 (Hz)} & \textbf{N1 (Hz)} & \textbf{FRR}  \\
        \cmidrule{2-4}\cmidrule{5-7}\cmidrule{8-10}
        FN class 0                                    & 164                                    & 14                                     & \textbf{11.7}                                  & 136         & 58          & 2.3          & 18          & 2           & 9.0          \\
        FN class 1                                    & 24                                     & 122                                    & 5.1                                   & 58          & 144         & \textbf{2.5}          & 0           & 36          & \textbf{$>$100}     \\
        TEST (mean)                            & 138                                    & 123                                    & 1.1                                   & 143         & 123         & 1.2          & 17          & 14          & 1.9          \\
        % TEST (max FRR)                         & 150                                    & 100                                    & 1.5                                   & 154         & 94          & 1.6          & 32          & 10          & 3.2          \\
        \botrule
    \end{tabular}
    \caption{
        \textbf{Output firing rates and FRRs for target and test input samples.}
        ``Simulated'' results are from PC-based simulation of the trained DynapSim network in Rockpool.
        ``Quantized'' results are from PC-based simulation of the quantized model in Rockpool.
        ``Hardware'' results are obtained from running inference of the model deployed to DYNAP-SE2 hardware.
        1000 samples were used for the Simulated and Quantized results.
        10 samples were used for inference on hardware.
        FN:~Frozen Noise (trained target input sample);
        TEST:~Random poisson untrained test samples;
        N0:~Output neuron for class~0;
        N1:~Output neuron for class~1.
    }
    \label{tab:fn_rate_response}
\end{table*}

We presented 1000 randomly generated independent test samples for inference in the simulation, taking advantage of the flexibility of the simulation environment. 
In contrast, for hardware testing, we presented 10 random independent Poisson noise samples to DYNAP-SE2. 
The limited number of hardware tests was a deliberate decision, influenced by the intricate nature of the hardware setup. 
Chip configuration for each iteration requires manual intervention, and required at least 5 minutes for each iteration in practice.

In all cases, mismatch was either simulated (for ``simulated'' and ``quantized'' results), or was physically present on the DYNAP-SE2 hardware device.
Untrained test samples produced high output firing rates in general (>100~Hz simulation; TEST in Tab.~\ref{tab:fn_rate_response}), but with low FRR close to 1.
The worst-case, highest FRR for the untrained test samples, indicating false-positive discrimination, was 1.5 in simulation; 1.6 for the quantized network; and 3.2 for inference on the DYNAP-SE2 HW.
Trained target frozen noise input samples produced similar maximum firing rates as the untrained test samples, but with much higher FRR.

The experimental results indicate that quantization and mismatch cause information loss when converting an optimized network to a hardware configuration.
However, despite this loss of information and device mismatch, the decision mechanism remains functional.
The hardware deployed model also successfully distinguishes trained input samples, with high FRR.

\paragraph{Training speed}

Gradient-based spiking neural networks training is known to take a long time due to the complexity of the dynamical equations that spiking neurons solve in time, which involve many floating point operations.
Additionally, backpropagation through time implies additional memory overhead compared with backpropagation in classical ANN optimisation problems.
However, recent advancements in machine learning tools have presented opportunities for performance improvements.

DynapSim utilizes JAX, a high-performance mathematical library which includes automatic differentiation.
JAX's just-in-time (JIT) compilation support for functional programming significantly reduces execution time in the optimization loop.
To illustrate the benefits of JIT, we executed the training script on two different machines, and the performance results are presented in Tab.~\ref{tab:fn_env}.

\begin{table}[t]
    \centering
    \begin{tabular}{@{}lll@{}}
        \toprule
        \textbf{Attribute} & \textbf{Machine 1}  & \textbf{Machine 2}   \\
        \midrule
        CPU                & 8 Core Apple M1 Pro & Intel Core i7-7500 \\
        RAM                & 32 GB               & 16 GB                \\
        OS                 & macOS 12.4          & Ubuntu 20.04         \\
        Epoch/s (JAX)      & 0.7                 & 0.4                  \\
        Epoch/s (JAX-JIT)  & 2600                & 1350                 \\
        Duration (JAX)     & 15 days             & 28 days              \\
        Duration (JAX-JIT) & 6.5 minutes         & 12.5 minutes         \\
        Speedup            & $3714\times$        & $3375\times$         \\
        \botrule
    \end{tabular}
    \caption{
        \textbf{Training time comparison.}
        The training process was executed identically using the same training code on two machines.
        We compared the training speed when using non-accelerated JAX, and when using JAX-JIT compilation to the CPU on each machine.
    }
    \label{tab:fn_env}
\end{table}

Using JAX-JIT reduced the computation time by up more than 3000$\times$, making it possible to optimize SNN structures employing complex neuron models without the need for giant computer clusters or waiting for weeks to see the results.
Rockpool / DynapSim is able to use the JIT facilities of JAX to target GPUs and TPUs, enabling scalable use of large computational resources when available, for efficient training of SNNs.

\section{Discussion}\label{conclusion}

We demonstrated a new approach and toolchain for gradient-based training and automated deployment of SNN applications to mixed-signal SNN devices such as DYNAP-SE2.
The training pipeline is shown to run 1 million epochs in minutes instead of weeks, by exploiting the just-in-time compilation features of JAX.
DynapSim is a huge step towards building commercial SNN applications for mixed-signal neuromorphic processors.

The deployment strategy offers an unsupervised method for weight quantization, removing the need for manual calibration and tuning of hardware bias parameters.
The resulting quantized network's parameters are automatically translated to a hardware configuration.
Although the task introduced here is relatively simple, it proposes a novel methodology for application development targeting mixed-signal SNN processors.
The approach, metrics, and evaluation strategies can easily be applied to more complex tasks.

Results indicate that the optimized network is robust to the effects of quantization and device mismatch, which are common challenges in hardware implementation. 
This suggests that spiking neural networks can be used robustly in real-world applications where hardware constraints and variability are significant factors.
Still, further studies are needed to investigate the network's performance under different quantization and device mismatch scenarios and to generalize the findings to different spiking neural network architectures and applications.

Rockpool simplifies the modeling and deployment process for the neuromorphic community, addressing a key obstacle that has limited the accessibility of mixed-signal neuromorphic processors to only high-end academic and industrial research.
Our approach paves the way for building commercial applications using mixed-signal neuromorphic technologies.

\section{Acknowledgements}

The authors thank Dmitrii Zendrikov and Adrian Whatley for their comments and feedback while developing the simulation tools.

This work was funded in part by the ECSEL Joint Undertaking under grant agreements 826655 ``TEMPO'' and 876925 ``ANDANTE''; by the KDT Joint Undertaking under grant agreement 101097300 ``EdgeAI''; by Innosuisse and by the Swiss State Secretariat for Education, Research and Innovation (SERI).

%%===========================================================================================%%
%% If you are submitting to one of the Nature Portfolio journals, using the eJP submission   %%
%% system, please include the references within the manuscript file itself. You may do this  %%
%% by copying the reference list from your .bbl file, paste it into the main manuscript .tex %%
%% file, and delete the associated \verb+\bibliography+ commands.                            %%
%%===========================================================================================%%

\bibliography{bibliography}

% \clearpage
\appendix
\section{Appendix}
\begin{table*}[!ht]
    \centering
    \footnotesize
    \begin{tabular}{@{}lll@{}}
        \toprule
        \textbf{Parameter} & \textbf{Corresponding current}  & \textbf{Description}   \\
        \midrule
        SOAD\_TAU\_P       & $I_{\tau_{ahp}}$   & AHP block time constant $\tau_{ahp}$                  \\
        DEAM\_ETAU\_P      & $I_{\tau_{ampa}}$  & Excitatory AMPA synapse time constant $\tau_{ampa}$   \\
        DEGA\_ITAU\_P      & $I_{\tau_{gaba}}$  & Inhibitory GABA synapse time constant $\tau_{gaba}$   \\
        DENM\_ETAU\_P      & $I_{\tau_{nmda}}$  & Excitatory NMDA synapse time constant $\tau_{nmda}$   \\
        DESC\_ITAU\_P      & $I_{\tau_{shunt}}$ & Inhibitory SHUNT synapse time constant $\tau_{shunt}$ \\
        SOIF\_LEAK\_N      & $I_{\tau_{mem}}$   & Neuron membrane time constant $\tau_{mem}$            \\
        SOAD\_PWTAU\_N     & $I_{pulse\_{ahp}}$ & AHP block pulse width $t_{pulse\_{ahp}}$     \\
        SYPD\_EXT\_N       & $I_{pulse}$       & Any synaptic input pulse width $t_{pulse}$  \\
        SOIF\_REFR\_N      & $I_{ref}$         & Neuron membrane refractory period $t_{ref}$ \\
        SOAD\_GAIN\_P      & $I_{gain_{ahp}}$   & AHP block gain                \\
        DEAM\_EGAIN\_P     & $I_{gain_{ampa}}$  & Excitatory AMPA synapse gain  \\
        DEGA\_IGAIN\_P     & $I_{gain_{gaba}}$  & Inhibitory GABA synapse gain  \\
        DENM\_EGAIN\_P     & $I_{gain_{nmda}}$  & Excitatory NMDA synapse gain  \\
        DESC\_IGAIN\_P     & $I_{gain_{shunt}}$ & Inhibitory SHUNT synapse gain \\
        SOIF\_GAIN\_N      & $I_{gain_{mem}}$   & Neuron membrane gain          \\
        SYAM\_W0\_P        & $I_{w_{0}}$      & Weight bit 0 strength             \\
        SYAM\_W1\_P        & $I_{w_{1}}$      & weight bit 1 strength             \\
        SYAM\_W2\_P        & $I_{w_{2}}$      & weight bit 2 strength             \\
        SYAM\_W3\_P        & $I_{w_{3}}$      & weight bit 3 strength             \\
        SOAD\_W\_N         & $I_{w_{ahp}}$    & AHP block weight current \\
        SOIF\_DC\_P        & $I_{dc}$         & Constant DC current injected as input \\
        DENM\_NMREV\_N     & $I_{if_{nmda}}$  & NMDA gate soft cut-off current        \\
        SOIF\_SPKTHR\_P    & $I_{spkthr}$     & spiking threshold current             \\
        \botrule
    \end{tabular}
    \caption{
        \textbf{Bias Parameters} A supplementary table listing the most significant parameters and their impact on the SNN simulation
    }
    \label{tab:all_params}
\end{table*}
\end{document}